\documentclass[preprint,pre]{revtex4-1}
\usepackage{graphicx}
\usepackage{amsmath}
\usepackage{amssymb}
\usepackage{natbib}
\begin{document}
 
\title{Anomalous Thermomechanical Properties of a Self-propelled Colloidial Fluid}

\author{S. A. Mallory$^1$, A.\ \v{S}ari\'c$^1$, C. Valeriani$^{2}$, A. Cacciuto$^1$}
\email{ac2822@columbia.edu}
\affiliation{$^1$Department of Chemistry, Columbia University\\ 3000 Broadway, New York, NY 10027\\ }
\affiliation{$^{2}$Departamento de Quimica Fisica, Facultad de Ciencias Quimicas, Universidad Complutense de Madrid, 28040 Madrid, Spain}
 
\begin{abstract}
We use numerical simulations to compute the equation of state of a suspension of spherical self-propelled nanoparticles in two and three dimensions. We study in detail the effect of excluded volume interactions and confinement as a function of the system's temperature, concentration and strength of the propulsion. We find a striking non-monotonic dependence of the pressure with the temperature, and provide simple scaling arguments to predict and explain the occurrence of such anomalous behavior. We conclude the paper by  explicitly showing how our results have important implications for the effective forces on passive components suspended in a bath of active particles.
\end{abstract}

%%\keywords{latex-community, revtex4, aps, papers}

\maketitle

The kinetic theory of gases proved to be one of the crowning achievements of $19^{th}$ century physics.  The seminal work of Bernoulli, Clausius, Maxwell  and Boltzmann presented a definitive relationship between the internal structure of a gas and its thermomechanical properties, and thus provided unprecedented insight into the behavior of gaseous systems and set the foundation for the  modern kinetic theory. 

In recent years, active systems have been at the forefront of non-equilibrium statistical mechanics, as 
they show a range of exotic behavior not typically observed in their passive counterparts, including turbulence \cite{Wolgemuth2008a,Wensink2012a,Wensink2012c,Giomi2012}, delayed crystallization \cite{Bialke2012}, and self-regulation \cite{Gopinath2012,Kaiser2012,Yang2012a,Angelani2011}. 
(see references \cite{Marchetti2012,Romanczuk2012,Ramaswamy2010a}
for recent reviews on the field). 
In spite of the great effort deployed  to systematically quantify the individual and collective dynamics of synthetic and naturally occurring active nanocomponents, such as self-propelled colloidal particles and bacterial suspensions,
our understanding of active systems still remains incomplete.

As an attempt to provide a better link between the microscopic properties and the resulting macroscopic behavior in actives systems, we consider a solution of spherical self-propelled nanoparticles (one of the simplest realizations of an active system) and determine what is typically one of the most fundamental properties of a solution: its equation of state. We show how self-propulsion  leads to anomalous thermomechanical properties, and how these are affected by confinement and excluded volume interactions. We also discuss, as an application of our results, how the effective interactions induced by an active ideal gas on two plates kept in close proximity present a new and unexpected  behavior as a function of activity and temperature.

To understand the interplay between active and thermal forces, we begin our study with what is possibly the simplest thermodynamic system: a dilute suspension of non-interacting self-propelled particles $-${\it the active ideal gas} $-$ and we ignore any effective interaction between the particles that may arise, for instance as a result of gradients in fuel concentration or  hydrodynamic interactions \cite{Brotto2013,Molina2013,Sanchez2012a,Schwarz-Linek2012}.
In our case, the pressure is determined solely by the average force that particles exert on an enclosing container.

We implement a minimal model, inspired from recent experimental and theoretical work \cite{Theurkauff, Palacci, Buttinoni, Redner, Stenhammar}, of $N$ self-propelled particles in two and three dimensions modeled as spheres of diameter $\sigma$ confined within a circular/spherical container of radius $R$ centered at the origin of our coordinate system.  Each ideal particle undergoes Langevin dynamics at a constant temperature $T$, and interacts exclusively with the wall via a truncated harmonic potential of the form
\begin{equation*}
V_w(r)=
\begin{cases}
   
0, & \text{if } r-R < 0 \\
k \left( r-R\right)^2, & \text{if } r-R \geq 0
\end{cases}
\end{equation*}

\noindent where $r$ is the distance from the center of the container, $k=800 k_{\rm B}T_0/\sigma^2$ is the spring constant, and $k_{\rm B}T_0$ (with $T_0=1$) is the thermal energy at room temperature that will be used as the energy scale.  

Self-propulsion is introduced through a directional force which has a constant magnitude, $|F_a|$,  and is directed along a predefined orientation vector, $\pmb{n}$,  which passes through the origin of each particle and connects its poles.  The equations of motion of an individual particle are given by the coupled Langevin equations

\begin{align}
m \ddot{ \pmb{r}}  & =-\gamma \dot{ \pmb{r}}-\partial_{\pmb{r}} V_w+|F_a|\pmb{n}+ \sqrt{2 \gamma^2 D} \pmb{\xi}(t)  \\
\dot{\pmb{n}} & = \sqrt{2D_r} \pmb{\xi_R(t)} \times \pmb{n}
\end{align}

\noindent where $m$ is the particle's mass, $\gamma$ is the  drag coefficient, and $D$ and $D_r$ are the translational and rotational diffusion constants, respectively.  The typical solvent induced Gaussian white noise terms for both the translational and rotational motion are characterized by $\langle \xi_i(t) \rangle = 0$ and   $\langle \xi_i(t) \cdot   \xi_j(t') \rangle = \delta_{ij}\delta(t-t')$ and $\langle  \xi_{Ri}(t)\rangle = 0$ and   
$\langle  \xi_{Ri}(t) \cdot   \xi_{Rj}(t') \rangle = \delta_{ij}\delta(t-t')$, respectively.  In the low Reynolds number regime the rotational and translation diffusion coefficients for a sphere satisfy the relation $D_r=(3D)/\sigma^2$.  
 
All simulations were carried out in 2 and 3 dimensions using LAMMPS~\cite{Plimpton1995}.
The mass of each particle is set to 1,   the drag coefficent $\gamma=10$, and the time step to $\Delta t=10^{-3}$.  The drag coefficient $\gamma$ was chosen to be sufficently large such that the motion of the particles is effectively overdamped.  Several of the simulations were repeated with larger values of $\gamma$ (e.g. $\gamma=50,100$), which produced no detectable differences in our results.   Each simulation was run for a  minimum of $3 \times10^7$ time steps.  The total number of particles ranged from $N=10^2$ to $N=3\times10^4$.  The mechanical pressure was calculated as the average forces the particles exerts on the boundary  divided by the  area of the container $A$. 

The zero-propulsion and zero-temperature  limits are well understood. In the former, the system reduces to an ideal gas exerting a pressure $P=\rho k_{\rm B}T$, while in the latter,   all particles  accumulate on the walls of the container and each contributes a persistent  force of magnitude $|F_a|$.  This result is due to a particle's inability to rotate in the absence of thermal fluctuations, and after hitting the wall they slide across its surface until the tangential component of the active force vanishes. The resulting pressure is trivially $P=N|F_a|/A$.

\begin{figure}[h!]
 \centering
\includegraphics{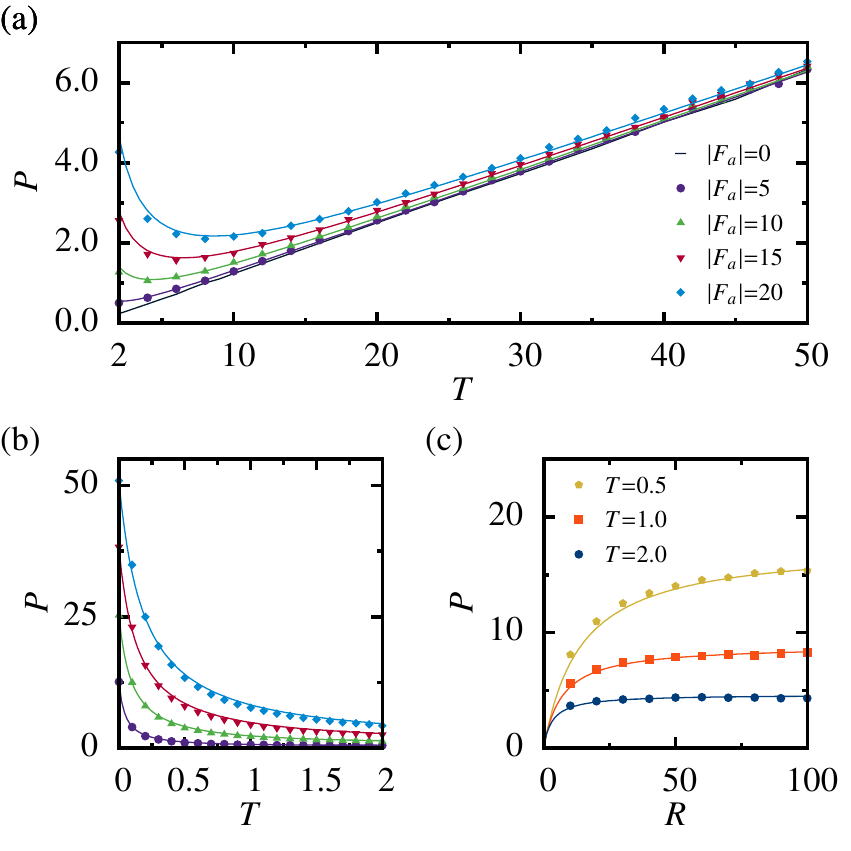}
  \caption{\label{fig:press_pvt2}(a) {Pressure in two-dimensional system as a function of  temperature for several values of $|F_a|$.  (b) The same curve as (a) for  $T<2$. (c) Pressure as a function of the box radius where $|F_a|=20$ for several values of $T$. All solid lines are plotted using Eq. \eqref{eq:last} and the corresponding system variables.}}
\end{figure}

\begin{figure}[h!]
 \centering
\includegraphics{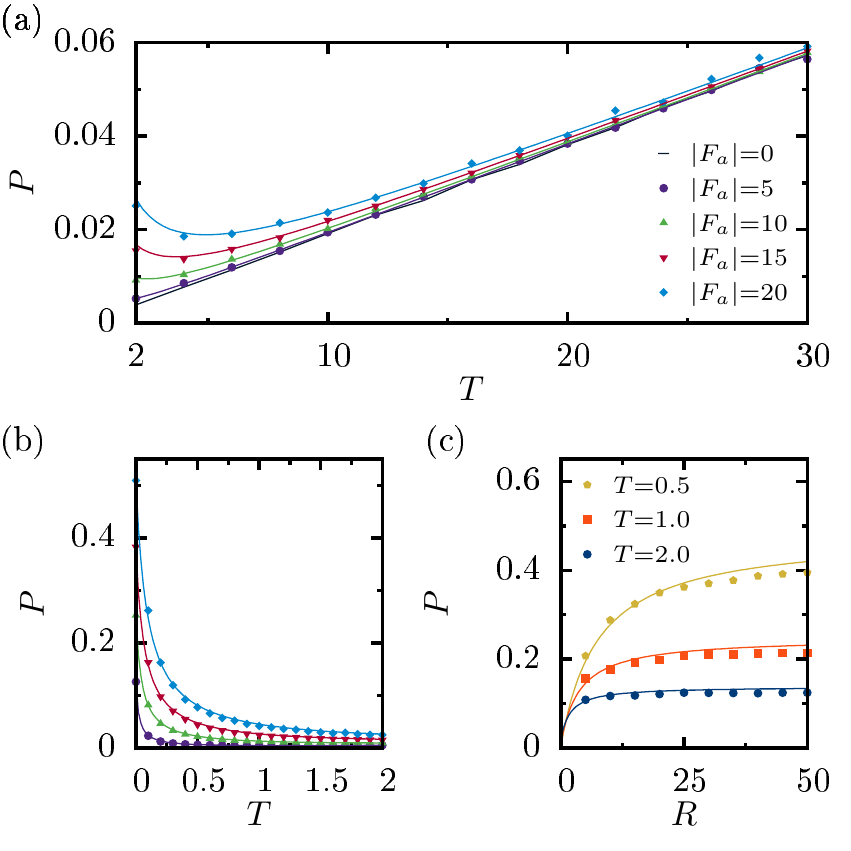}
\caption{\label{fig:press_pvt3}(a) {Pressure in three-dimensional system as a function of  temperature for several values of $|F_a|$.  (b) The same curve as (a) for  $T<2$. (c) Pressure as a function of the box radius where $|F_a|=20$ for several values of $T$. All solid lines are plotted using Eq. \eqref{eq:last} and the corresponding system variables.}} 

\end{figure}
 
The interesting behavior emerges at finite temperatures and moderate activity where the rotational diffusion is able to derail the otherwise rectilinear trajectories of the particles.  Figures \ref{fig:press_pvt2}(a) and \ref{fig:press_pvt3}(a) show how the pressure exerted on the wall depends on the temperature $T$ for different values of the active force respectively in two and three dimensions. Strikingly, these curves show non-monotonic behavior. Specifically, for small values of $T$, an increase of the temperature leads to a sharp decrease of the pressure, whereas for large values of $T$,  the system's pressure re-establishes the  ideal gas pressure-temperature dependence expected for non-active particles. 

The behavior of the pressure at low temperatures (or large $|F_a|$) can be rationalized with the following argument:
The two relevant time scales in the system are the decay of the rotational correlation time due to thermal fluctuations, 
this can be written for the specific cases of two and three dimensions as $\tau_r \simeq (d-1)D_r^{-1}$ \cite{Hagen2009}, where $d=2,3$ is the system dimensionality, and 
the time required for the particles to move across the system.
The latter time-scale at low $T$ is dominated by the active motion, and scales as $\tau_a\simeq R\gamma/|F_a|$. Therefore, as long as  
$\tau_a/\tau_r\ll1$,  i.e. the rotational motion does not decorrelate faster than the time required for the particle to cross the container, there will always be a net linear contribution to the pressure on the wall by the active force.  
A simple way of estimating the average pressure in this regime can be  obtained by 
considering that a particle can be either on the wall pushing with a force proportional to $|F_a|$, or diffusing across the box generating no pressure on the wall.
A particle on the wall will exert a force for a time  of  the order of $\tau_r\simeq (d-1)D_r^{-1}$. The average time spent by the particle diffusing across the box  without generating any pressure on the wall is  $\tau_a\simeq R\gamma/|F_a|$.
Hence, over the time $\tau_r+\tau_a$ the force exerted on the wall by a single active particle can be estimated  as
$\langle F \rangle \simeq (|F_a| \cdot \tau_r + 0 \cdot \tau_a)/(\tau_r + \tau_a)= |F_a|/(1+\tau_a/\tau_r)$.

Multiplying by $N$ and dividing by the surface, one obtains the total pressure, 
\begin{equation}
\langle P \rangle = \left(\frac{N}{A}\right )\frac{|F_a|}{1+\frac{\tau_a}{\tau_r}} \,, 
\label{thiss}
\end{equation}
and in the limit  $\tau_a/\tau_r \ll1$,  this can be further simplified to 
 \begin{equation}
\langle P \rangle = \left(\frac{N}{A}\right )\frac{|F_a|}{1+\frac{\tau_a}{\tau_r}} \simeq \frac{N|F_a|}{A}(1-\frac{\tau_a}{\tau_r})\simeq \frac{N|F_a|}{A}e^{- \frac{\tau_a}{\tau_r}}
  \label{this2}
 \end{equation}

We expect {that the exact  numerical value of the pressure will be dependent on the specific geometry of the boundary, as the behavior of the particles on the surface is well known to be strongly sensitive to it
\cite{gompper}

, but the qualitative nonmonotonic behavior observed here should remain unaltered.}  
To account for this uncertainty, we introduce a geometric factor $\alpha$ and Eq.~\eqref{thiss} can be recast as 
\begin{equation}
\langle P \rangle = \left(\frac{N}{A}\right )\frac{|F_a|}{1+\frac{\tau_a}{\tau_r}}=\left(\frac{R}{d}\right) \frac{\rho|F_a|}{1+\alpha  \left (\frac{3(d-1)Rk_{\rm B}T}{\sigma^2|F_a|}\right )}
  \label{this3}
 \end{equation}
where we have also used the appropriate Einstein relation $(D=k_BT/\gamma)$ to introduce the proper temperature dependence.

\begin{figure}[h!]
  \centering
   \includegraphics{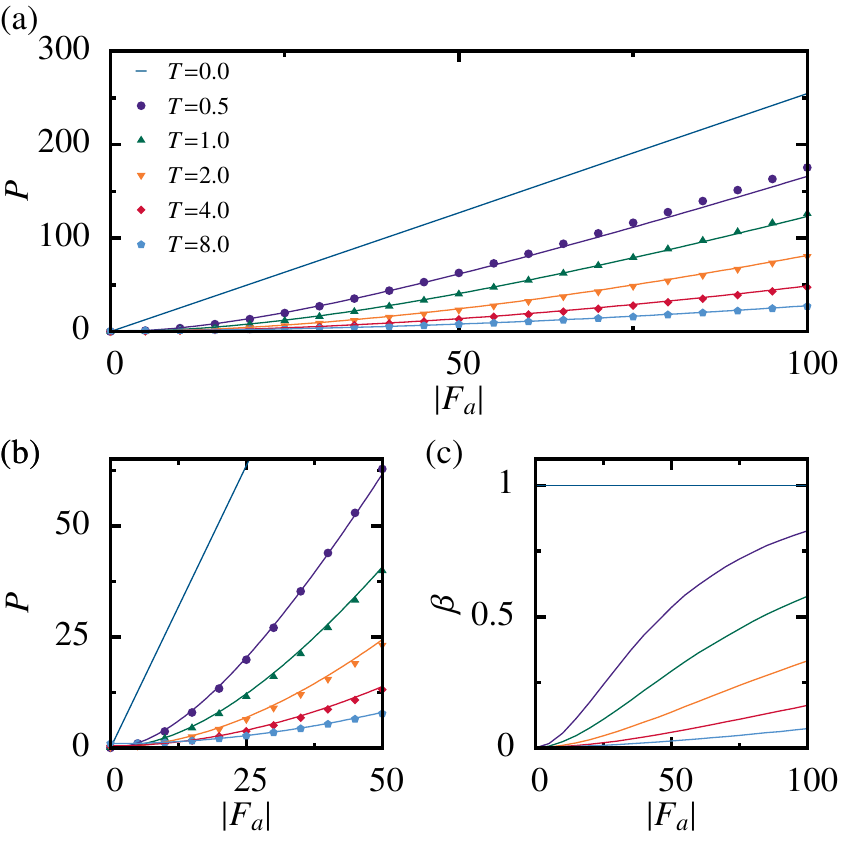}
    \caption{\label{fig:press_pvk2} (a) Pressure in two-dimensional system as a function of $|F_a|$ for several values of $T$.  (b) The same curve as (a) for  $|F_a|<50$.  The solid lines in (a) and (b) correspond to Eq. \eqref{eq:last}.  (c)  The fraction of particles, $\beta$, that are within one particle radius from the boundary as a function of  $|F_a|$ for various values of $T$. }
\end{figure}

\begin{figure}[h!]
  \centering
   \includegraphics{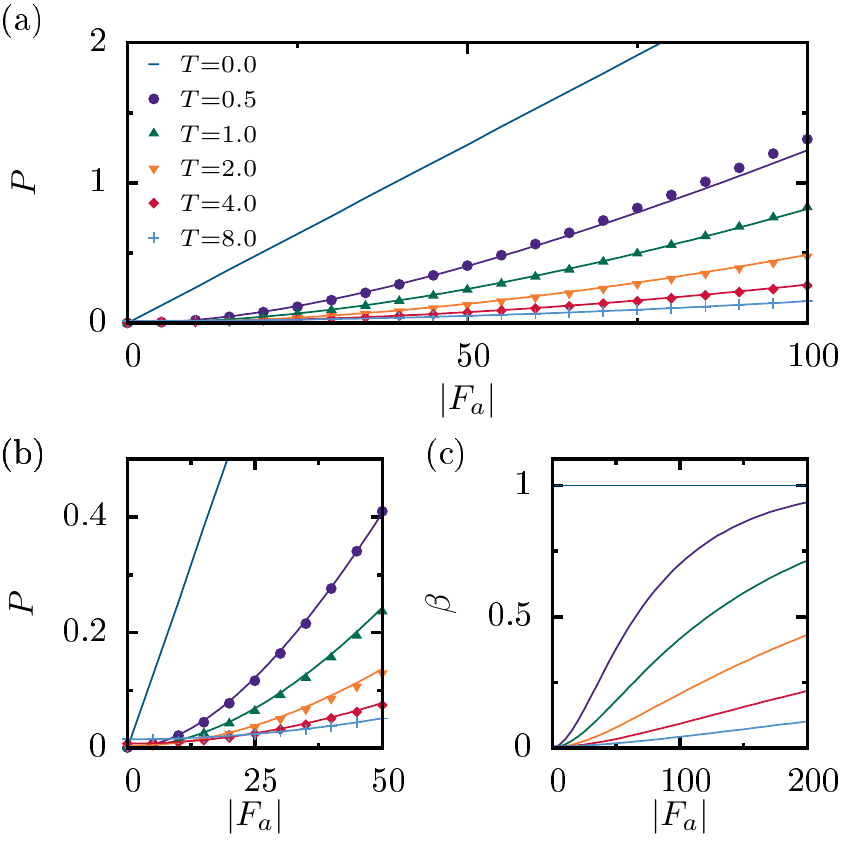}
  \caption{\label{fig:press_pvk3}  Pressure in three-dimensional system as a function of $|F_a|$ for several values of $T$.  (b) The same curve as (a) for  $|F_a|<50$.  The solid lines in (a) and (b) correspond to Eq. \eqref{eq:last}.  (c)  The fraction of particles, $\beta$, that are within one particle radius from the boundary as a function of  $|F_a|$ for various values of $T$. }

\end{figure}

The derivation of Eq. \eqref{this3} assumes that particles spend a majority of their time at the boundary, which ceases to be true when $\tau_a/\tau_r  \gg 1$. A simple expansion of Eq.~\eqref{this3} in this limit gives
$\langle P \rangle \simeq \left(\frac{\rho}{d}\right ) \frac{|F_a|^2\sigma^2}{3(d-1)k_{\rm B}T}$ indicating that, as is typical for nonactive ideal gases,
the pressure depends on the square of the particle velocities. Of course, such an expansion assumes that the 
particle velocities are proportional to the active force, but at large temperatures, one cannot neglect the contribution of thermal fluctuations to the motion of the particle.  In this regime, we appeal to the insights of equilibrium statistical mechanics and make the assumption that the pressure is proportional to the average translational kinetic energy of the system.  The introduction of self-propulsion when  $\tau_a/\tau_r  \gg 1$ can be viewed as a  perturbation of an ideal gas.  One way to associate a pressure in this regime is to compute $\langle \dot{\pmb{r}}^2(t)\rangle $ from Eq. (1) (See Appendix A).  For simplicity, we consider particles confined to two dimensions and only have a single degree of rotational freedom.  Assuming the system reaches its steady state (i.e. $t \rightarrow \infty$ and $\langle F \rangle =0$),  we can write
\begin{equation}
\begin{split}
\langle \dot{\pmb{r}}(t)^2 \rangle = \langle \dot{x}(t)^2 \rangle + \langle \dot{y}(t)^2 \rangle = & \frac{2k_BT}{m}
+\frac{|F_a|^2}{ m (3k_BT/\sigma^2+\gamma^2/m)}
\end{split}\label{v2}
\end{equation}
Measurements of $\langle \dot{\pmb{r}}(t)^2 \rangle$ in numerical simulations performed in two dimensions on a system with periodic boundary conditions are in good agreement with Eq.~\ref{v2}.

Assuming that in this limit  we can write the pressure as 
$P/\rho=\frac{m}{2}\langle \dot{r}(t)^2 \rangle$, 
we obtain
\begin{equation}
P=\rho \left (k_{\rm B}T+\frac{ |F_a|^2}{6k_BT/\sigma^2+2\gamma^2/m}\right )
\label{eq:P_low_Pe}
\end{equation}
\noindent 
where $\rho$ is the number density. The first term captures the ideal gas behavior, whereas the second term corrects for the increase of the velocity of the particles due to their propulsion. This correction is not a constant factor, but slowly decays with the temperature as the extent of the random thermal forces overwhelms (affecting both rotational and linear degrees of freedom) the role of the active ones.  It is important to stress that, if the pressure/velocity relation that we used in Eq. \eqref{eq:P_low_Pe} is indeed applicable to an active gas, it can only be  valid for very high temperatures and large values of $R$.

Remarkably, we find that the equation obtained by simply adding the ideal gas limit $\rho k_{\rm B}T$ to Eq.~\ref{this3}, 
\begin{equation}
P=\rho k_{\rm B}T+\left(\frac{R}{d}\right) \frac{\rho|F_a|}{1+\alpha  \left (\frac{3(d-1)Rk_{\rm B}T}{\sigma^2|F_a|}\right )}
\label{eq:last}
\end{equation}
provides the best description of the data observed in our simulations in both two and three dimensional systems in terms of all external parameters: the temperature (Fig.~\ref{fig:press_pvt2} and Fig.~\ref{fig:press_pvt3} panels (a) and (b)), the radius of the cavity (Fig.~\ref{fig:press_pvt2} and Fig.~\ref{fig:press_pvt3}  panels (c)), and the strength of the activity (Figs.~\ref{fig:press_pvk2} and Fig.~\ref{fig:press_pvk3} panels (a) and (b)).    The single fitting parameter for all curves in two and three dimensions is estimated to be $\alpha\simeq 0.89$.  
{Our derivations of Eqs.~\ref{this2} and ~\ref{this3} is based on simple scaling arguments, however, a recent detailed analysis of the dynamics of self-propelled particles under strong confinement near the boundary has been carried out by Fily et al.\cite{Fily}. They found the analogous exponential decay we report in Eq.~\ref{this2}. In their calculations they predict that $\alpha=1$, which is in good agreement with our numerical findings.}
The (c) panels on Fig.~\ref{fig:press_pvk2} and Fig.~\ref{fig:press_pvk3} also indicate how the average fraction of particles on the surface of a container, $\beta$, defined as the number of particles that are within a particle radius, $\sigma/2$, from the boundary divided by the total number of particles in the system, depends on  $|F_a|$, for different values of the temperature. At low temperatures/large-propulsions the particles tend to accumulate at the surface, while for large-temperatures/weak-propulsions the particles are dispersed homogeneously  within the container.

It should again be stressed that Eq.~\eqref{eq:last} suggests that the anomalous behavior of the pressure with the temperature is not solely to be found  in highly confined gases, but should persist in the limit of large containers.

 Taking the limit for $R\rightarrow\infty$ in Eq.~\eqref{eq:last} 
while keeping the density constant gives 
\begin{equation}
P=\rho \left ( k_{\rm B}T+ \frac{|F_a|^2\sigma^2}{3\alpha d(d-1)k_{\rm B}T}\right)  \,\,\,\,\,\,\,\, d=2,3
\end{equation}
which has the same limiting behavior for very high temperatures as Eq.~\ref{eq:last}.
The location of the inflection point, $T^*$, whose value grows linearly with the strength of the active force
and is independent of $R$,  can then be easily computed to  give
\begin{equation}
T^*=\frac{|F_a|\sigma}{k_{\rm B} \sqrt{3\alpha d(d-1)}} \,\,
\end{equation}

Let's now consider the behavior of an active gas when excluded volume interactions between the particles are no longer negligible. This is achieved in our simulations via the Weeks Chandler Andersen (WCA) potential 
\begin{equation}
 U(r_{ij})=4 \epsilon \left[ \left( \frac{\sigma}{r_{ij}} \right)^{12}- \left( \frac{\sigma}{r_{ij}} \right)^{6}+\frac{1}{4} \right]  
  \label{eq:LJ_V}
\end{equation} 
\noindent with a range of action  extending up to $r_{ij}=2^{1/6}\sigma$. Here $r_{ij}$ is the center to center distance between any two particles $i$ and $j$, and we set $\epsilon=10k_{\rm{B}}T_0$. 

\begin{figure}[b!]
  \centering
 \includegraphics{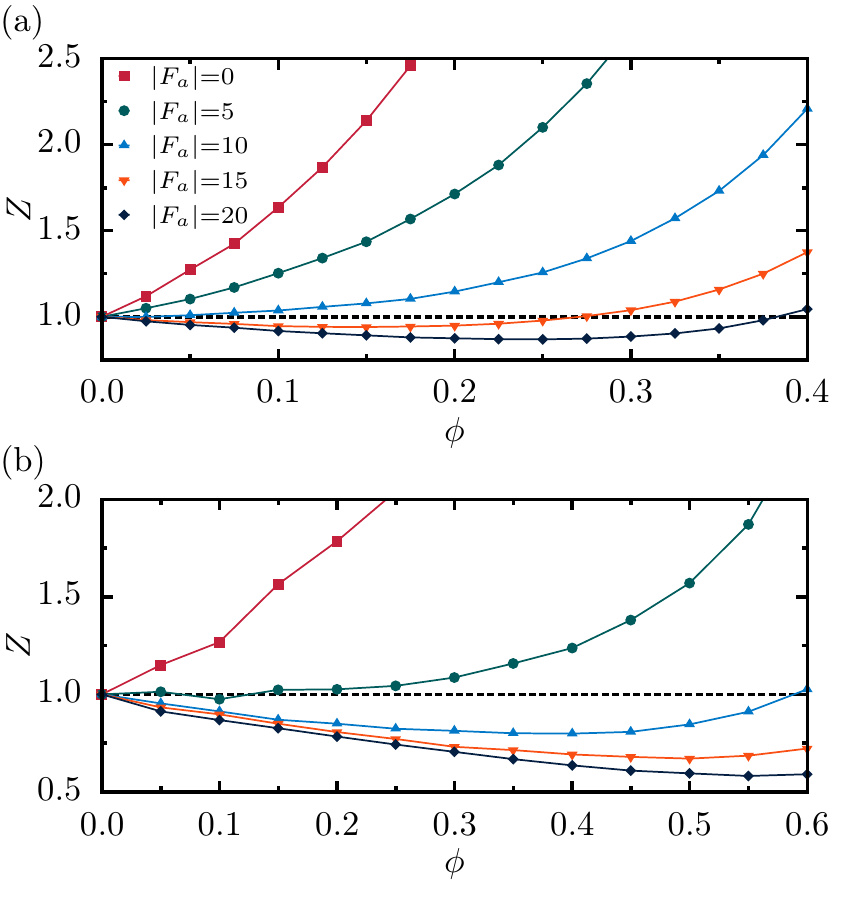}
  \caption{\label{fig:press_pvd}Compressibility  factor, $Z$, for three (a) and two (b) dimensional systems as a function of volume fraction, $\phi$, for different values of the active force, $|{F}_a|$, at $T=1$. The dashed line indicates the non-active ideal case.}
\end{figure} 

In analogy with non-active gases, the deviations from the ideal (non-interacting) case are estimated by $Z\equiv P/P_0$, where $P$ and $P_0$ are, respectively, the pressures of the non-ideal and ideal active systems at the same $V$, $T$ and $|F_a|$. The conventional understanding of this factor for non-active systems is intimately linked to the virial expansion of the pressure at low densities: when  $Z \approx 1$,  the system has nearly ideal behavior and molecular interactions are negligible, $Z>1$ is the  signature of a positive second virial coefficient pointing to an effectively repulsive nature of the interactions between the particles, and $Z<1$ would point to an effective attraction  that under the appropriate conditions would lead to phase separation. The compressibility factor as a function of volume fraction $\phi=\pi\rho\sigma^3/6$ at several values of $|F_a|$ is shown in Fig.~\ref{fig:press_pvd}.  As expected, for small values of $|F_a|$, the purely repulsive nature of the interparticle interactions leads to $Z>1$, for the range of volume fractions considered in these simulations. Strikingly, as the activity is increased, $Z$ shows non-monotonic behavior. Specifically, there exist  values of $Z$ smaller than one which appear for a range of low densities until excluded volume interactions eventually take over at larger 
$\phi$  to restore the expected $Z>1$. It is tempting to associate the value of $Z<1$ with an
effective attraction among the  particles leading to a phase separation between the high density layers on the surface and the low density in the bulk, however, it is not clear that a simple relation
between the virial coefficients and the sign of the interaction between the particles also holds for active non-ideal gases.
Nevertheless, phase separation of spherical self-propelled particles has been recently observed  and predicted
in two and three dimensional systems with periodic boundary conditions\cite{sp1,sp2,Redner,sp4,sp5,sp6}.

\begin{figure}[h!] 
  \centering
 \includegraphics{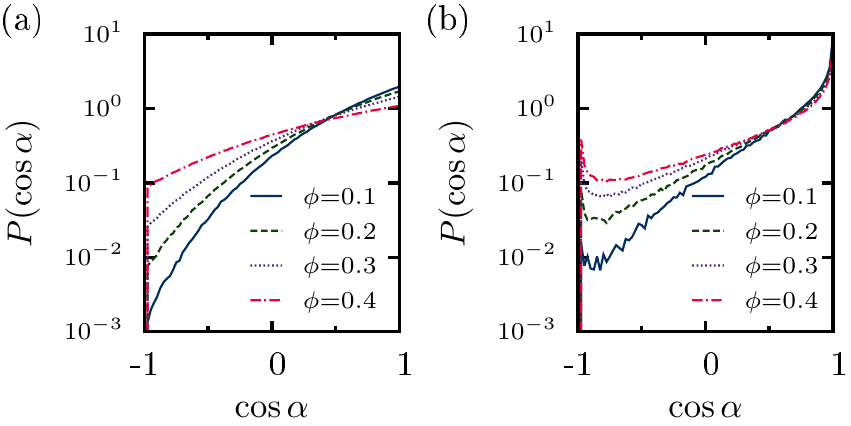}
  \caption{\label{fig:thetas2d} Distribution of the dot products between the propelling axis of a particle and the normal to the wall at the particle's location for various volume fractions of interacting particles in both three (a) and two (b) dimensions. Only  particles within a range of $\sigma$ from the wall are considered for this analysis. The log scale in the vertical axis has been chosen to better highlight the difference between the curves.  The value of the active force is $|F_a|=20$.}
\end{figure}

To understand this behavior it is instructive to consider how the active force (the dominant term in the large $|F_a|$ limit) generates a pressure on the wall.  For an ideal active gas,  particles accumulating at the walls  can freely slide across its surface until they align their propulsion axis parallel to the normal of the wall, thus each contributing the maximal force $|F_a|$(when $T=0$) to the pressure.  For a non-ideal active gas, excluded volume interactions between the particles hinder the particles' ability to reach these optimal configurations.  Clearly, this becomes more of a problem as the volume fraction is increased, and the particles begin to form layers at the surface of the container.
 The net result is a smaller average force per particle relative to the one generated by the ideal gas and a compressibility factor $Z<1$. This effect is highlighted in Fig.~\ref{fig:thetas2d} where we plot the angular distribution $P(\cos \alpha)$  for all the particles in contact with the boundary for various volume fraction in two and three dimensions.   As density increases, the angular distribution broadens, and a fraction of the particles are anti-aligned to the normal at the boundary, indicating that the second layers of particles can cage particles in the first layer leading to a broadening of the angular distribution.

\begin{figure}[t!] 
  \centering
  \includegraphics{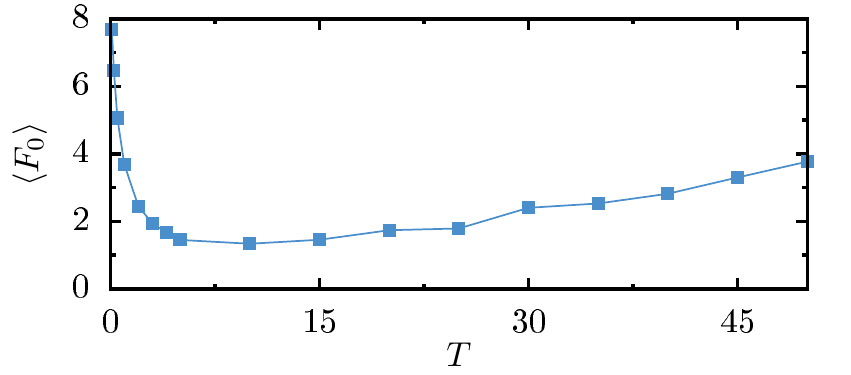}
  \caption{\label{fig:depletion} Non-monotonic behavior of the effective force, $\langle F_0\rangle$, induced by an active  gas on two plates held at fixed contact distance as a function of   $T$.  The value of the active force is $|F_a|=25$.}
\end{figure}

We briefly conclude with an important application of the nontrivial thermomechanical properties of an active gas reported in this paper. We compute the strength of the effective force felt by two parallel plates immersed in a two dimensional suspension of non-ideal active particles. For the sake of simplicity, we consider a two-dimensional system, where each plate has a length of $10 \sigma$ and a width of $\sigma$, and are fixed at a center to center distance $\sigma$ such that the faces of the plates are in contact.  All particles in the system have a diameter of $\sigma$ and interact via the potential in Eq.~\eqref{eq:LJ_V} with $\varepsilon=10\,k_{\rm B}T_0$.  The plates are made to be purely repulsive and interact with the particles in the system via a WCA potential with cutoff, $\sigma_R$, where $\sigma_R$ is the shortest distance between the center of the particle and the edge of the plate.  The strength factor is kept consistent with the interparticle strength factor and is set to $\varepsilon_R=10\,k_{\rm B}T_0$. In addition, the simulation box is setup to be a square of area $A$ with periodic boundary conditions at a fixed particle concentration $\rho=N/A=0.1$ with $N=188$ particles. The strength of the effective attractive force between the plates, $\langle F_0\rangle$ is computed  from the mean force acting on the plates when held in place.
The well known result for non-active systems is that the strength of the induced attractive potential (depletion force) is proportional to the local pressure imbalance $~\rho k_{\rm B}T$ (in the ideal gas approximation) that develops when the plates are at a surface-to-surface distance smaller than $\sigma$.  The expectation for non-active particles is therefore that the interaction strength should increase monotonically with  $T$ and $\rho$. Fig. \ref{fig:depletion} shows how instead the strength of the force between the plates presents a non-monotonic behavior with  temperature when the depletant is activated by a sufficiently large propelling force, $|{F_a}|=25$ in our two dimensional simulations. This counterintuitive result becomes immediately obvious when considering the anomalous pressure/temperature dependence discussed in this paper. We anticipate that other anomalies will appear when computing the full potential of mean force between the plates as  a function of distance, as also observed by Angelani et at.\cite{Angelani2008} for the case of active rod-like particles and spherical colloids.

\section*{Acknowledgments}
We thank Clarion Tung and Joseph Harder for insightful discussions and helpful comments. AC acknowledges financial supported from the National Science Foundation under CAREER Grant No. DMR-0846426. CV acknowledges financial support from a Juan de la Cierva Fellowship, from the Marie Curie Integration Grant PCIG-GA-2011-303941 ANISOKINEQ, and from the National Project FIS2013-43209-P.

\appendix 

\section{}

In the high temperature limit where  $\tau_a/\tau_r  \gg 1$, the average pressure exerted by a single active particle can be readily computed using two different methods.   For simplicity, we consider a particle confined to two dimensions with a single rotational degree  of freedom.  The first approach is simply to assume the pressure is proportional to the average squared velocity of the particle and thus $P/\rho=m\langle \pmb{\dot{r}}^2 \rangle/2$ where $\rho$ is the number density.  A more rigorous approach is to directly compute the correlation function $\langle \pmb{r} \cdot \pmb{F}_{e} \rangle$ where $\pmb{F}_e$ is the net force exerted on the particle. For a system of ideal particles, it is relatively straight forward to prove $P/\rho=- \langle \pmb{r} \cdot \pmb{F}_{e} \rangle$. \cite{Pathria}  In the long time limit, both approaches give the same expression for the pressure, and thus we choose to illustate the former method for its computational convienience and simple interpretation.  The equation of motion of an active Brownian particle of mass $m$ and self-propelling force $|F_a(\theta(t))|$ are given by the coupled Langevin Equations:
\begin{align}
m \ddot{ \pmb{r}}  & =-\gamma \dot{ \pmb{r}}+|F_a|\pmb{n}+\sqrt{2D} \pmb{\xi}(t)  \\
 \dot{\theta}& =\sqrt{2 D_r} \,\,\xi_R(t) 
\end{align}

Using a similar approach to that given in \cite{Hagen2009} and references therein, the x-component of the velocity is given by
\begin{equation}
\dot{x}(t)=\dot{x}_0e^{-\frac{\gamma}{m} t}+ \frac{e^{-\frac{\gamma}{m}t}}{m} \int^t_0\left[  |F_a|\cos(\theta(s))+\sqrt{2D}\xi(s)\right]e^{\frac{\gamma}{m} s} ds
\end{equation}

\noindent where $\dot{x}_0=\dot{x}(0)$. It follows that 
\begin{equation}
\begin{split}
\dot{x}(t)^2 &=\dot{x}_0^2e^{-\frac{2\gamma}{m} t}+ \frac{2\dot{x}_0e^{-\frac{\gamma}{m}t}}{m} \int^t_0 
\left[|F_a|\cos(\theta(s))+\sqrt{2D}\xi(s) \right]e^{\frac{\gamma}{m} s} ds \\
&+\frac{e^{-\frac{2\gamma}{m} t}}{m^2}\left[\int^t_0[  |F_a|\cos(\theta(s))+\sqrt{2D}\xi(s)]e^{\frac{\gamma}{m} s} ds\right]^2
\end{split}
\end{equation}

\noindent which readily simplifies to
\begin{equation}
\begin{split}
\dot{x}(t)^2 &=  \dot{x}_0^2e^{-\frac{2\gamma}{m} t} + \frac{2\dot{x}_0e^{-\frac{\gamma}{m}t}}{m} \int^t_0 \left[|F_a|\cos(\theta(s))+\sqrt{2D}\xi(s) \right]e^{\frac{\gamma}{m} s} ds \\ 
&+\frac{|F_a|^2e^{-\frac{2\gamma}{m} t}}{m^2}\int^t_0\int^{t}_0  \cos(\theta(s))  \cos(\theta(s'))e^{\frac{\gamma}{m} (s+s')} ds'ds   \\
&+\frac{2De^{-\frac{2\gamma}{m} t}}{m^2}\int^t_0\int^{t}_0\xi(s)\xi(s')e^{\frac{\gamma}{m} (s+s')} ds'ds
\end{split}
\end{equation}

\noindent Using the equipartition theorem ($\langle \dot{x}_0^2 \rangle=k_BT/m)$) and taking the ensemble average of $ \dot{x}(t)^2$ we are able to further simplify the above expression to
\begin{equation}
\begin{split}
\langle \dot{x}(t)^2 \rangle & = \frac{k_BT}{m}e^{-\frac{2\gamma}{m} t}  +\frac{2De^{-\frac{2\gamma}{m} t}}{m^2}\int^t_0 e^{\frac{2\gamma}{m}s} ds \\
&+ \frac{|F_a|^2e^{-\frac{2\gamma}{m}t }}{m^2}\int^t_0\int^{t}_0\langle  \cos(\theta(s))  \cos(\theta(s'))\rangle \,e^{\frac{\gamma}{m} (s+s')} ds'ds  
\end{split}
\label{equa}
\end{equation}

\noindent To evaluate the integral in the last term, it is important to note that $\langle \cos(\theta(t))  \cos(\theta(t'))\rangle_{t>t'} =e^{-D_r(t-t')}/2 $.  It follows that
\begin{equation}
\begin{split}
& \frac{|F_a|^2e^{-\frac{2\gamma}{m} t}}{m^2}\int^t_0\int^{t}_0 \langle \cos(\theta(s))  \cos(\theta(s'))\rangle e^{\frac{\gamma}{m} (s+s')} ds'ds \\
 & = \frac{|F_a|^2e^{-\frac{2\gamma}{m} t}}{m^2} \left( \int^t_0\int^{s}_0 \langle \cos(\theta(s))  \cos(\theta(s'))\rangle e^{\frac{\gamma}{m} (s+s')} ds'ds + \int^t_0\int^{t}_s \langle \cos(\theta(s))  \cos(\theta(s'))\rangle e^{\frac{\gamma}{m} (s+s')} ds'ds \right) \\
&=\frac{|F_a|^2e^{-\frac{2\gamma}{m} t}}{2m^2} \left( \int^{t}_0  e^{-(D_r-\frac{\gamma}{m})s} \left[\int^{s}_0 e^{(D_r+\frac{\gamma}{m}) s'}ds'\right]ds + \int^{t}_0  e^{(D_r+\frac{\gamma}{m})s} \left[\int^{t}_s e^{-(D_r-\frac{\gamma}{m}) s'}ds'\right]ds   \right) \\
& = \frac{|F_a|^2e^{-\frac{2\gamma}{m} t}}{2m^2} \left( \frac{1}{D_r+\frac{\gamma}{m}} \left[\frac{m}{2\gamma}\left[e^{\frac{2\gamma}{m}t}-1\right]+\frac{1}{D_r-\frac{\gamma}{m}}\left[e^{-(D_r-\frac{\gamma}{m})t}-1\right]\right] \right. \\
&\hspace{3.0 cm} +   \left. \frac{1}{D_r-\frac{\gamma}{m}}\left[\frac{m}{2\gamma}\left[e^{\frac{2\gamma}{m}t}-1\right]+\frac{1}{D_r+\frac{\gamma}{m}}\left[e^{\frac{2\gamma}{m}t}-e^{-(D_r+\frac{\gamma}{m})t}\right]\right] \right)
\end{split}
\end{equation}

The integral in the second term of Eq.~\ref{equa} is straightforward  and the final result is  
 
\begin{equation}
\begin{split}
\langle \dot{x}(t)^2 \rangle = & \frac{k_BT}{m} 
+\frac{|F_a|^2}{2m^2} \left(\frac{1}{D_r+\alpha} \left[\frac{1}{2\alpha}\left[1-e^{-2\alpha t}\right]+\frac{1}{D_r-\alpha}\left[e^{-(D_r+\alpha)t}-e^{-2\alpha t}\right] \right]\right. \\
& \hspace{4.0 cm} + \left. \frac{1}{D_r-\alpha} \left[\frac{1}{2\alpha}\left[1-e^{-2\alpha t}\right]-\frac{1}{D_r+\alpha}\left[1-e^{-(D_r+\alpha)t}\right] \right] \right)
\end{split}
\end{equation}

\noindent where $\alpha\equiv\frac{\gamma}{m}$.  In the limit that $t \rightarrow \infty$,
\begin{equation}
\begin{split}
\langle \dot{x}(t)^2 \rangle = & \frac{k_BT}{m}
+\frac{|F_a|^2}{2 m^2 \alpha (D_r+\alpha)}
\end{split}
\end{equation}

\noindent  A similar computations can be carried out to show that $\langle \dot{y}(t)^2 \rangle =\langle \dot{x}(t)^2 \rangle$ and thus the pressure is given by
\begin{equation}
\begin{split}
P/\rho &=\frac{m}{2}\langle \dot{r}(t)^2 \rangle = \frac{m}{2} \left( \langle \dot{x}(t)^2 \rangle + \langle \dot{y}(t)^2 \rangle \right) \\
&=m\langle \dot{x}(t)^2 \rangle  = k_BT+\frac{|F_a|^2}{2 \gamma (D_r+\gamma/m)} \\
&=k_BT+\frac{|F_a|^2}{6 k_BT/\sigma^2+2 \gamma^2/m}
\end{split}
\end{equation}

\end{document}